\documentstyle[12pt]{article}

\begin{document}

\title{Does the radioactive decay obey the Poisson statistics?}
\author{A. A. Kirillov \\
{\em Institute for Applied Mathematics and Cybernetics,}\\
{\em 10 Ulyanova str., } {\em Nizhny Novgorod, 603005, Russia}\\
e-mail: kirillov@unn.ac.ru}
\date{}
\maketitle

\begin{abstract}
It is shown that a nontrivial quantum structure of our space at macroscopic
scales, which may exist as a relic of quantum gravity processes in the early
universe, gives rise to a new fundamental phenomenon: spontaneous origin of
an interference picture in every physical process. This explains why
statistical distributions in radioactivity measurements may be different
from the Poisson distribution.
\end{abstract}

In this letter I would like to draw attention to the strange phenomenon
which was claimed to be observed in radioactivity measurements \cite{shn98}.
The phenomenon pointed out represents the fact that an instant shape of the
probability density distribution for the number of fissions, which should
obey the conventional Poisson distribution, apparently exhibits the
existence of a fine structure\footnote{
The claims of Ref. \cite{shn98} are more ambitious. However, from our point
of view the most important fact, which can be extracted from this work, is
the possible violation of the Poisson statistics.}. Presumably, this
structure evolves and disappears after averaging over some period of time.
The last fact would explain why it is difficult to observe this structure in
consecutive measurements (as in the case of radioactive decay) and why in
Ref. \cite{shn98} it was used a rather nonstandard procedure to analyze
measurement data.

The oddity (from the modern physics standpoint) of such a phenomenon causes
strong doubts in the existence of the effect itself and makes people to
think about artifacts or mistakes in the procedure of analyzing data (e.g.,
see Ref. \cite{der2000}). Doubts are supported, in the first place, by the
absence of any theoretical scheme which could explain the mechanism of the
origin and properties of this structure.

In this paper we show that such a mechanism really does exist. We suggest
one as an indirect result of quantum gravity effects, though this is surely
not the only possibility. It suffice to recall that the environment and the
structure of the radioactive sample itself should also leave an imprint on
the shape of the probabilistic distribution (e.g., simular effect exists in
energy $\beta -$decay spectra \cite{GF99}). Besides, there exists the
standard and more direct way to verify the existence of the fine structure
which (unlike the method used in Ref. \cite{shn98}) can be used by any
experimental group.

Now, let us recall the method of observing and analyzing the fine structure
used in Ref. \cite{shn98} and explain why we suppose the results to be
credible. First, one obtains long enough time series of measured signal,
e.g., a set of numbers of fissions $n$ during an interval $\ \Delta t$,
where $\ \Delta t$ is the shortest period of the measurement. Then, one
takes portions of this signal with a length $T$ $\gg \Delta t$ and
constructs a set of histograms $P_{j}\left( n\right) =N\left( n\right) /N$
(where $j=t/T$, $N\left( n\right) $ is the number of intervals $\Delta t$ in
which the number of fissions has the value $n$, and $N$ $=T/\ \Delta t$ is
the total number of intervals). This construction presumes that the function 
$P\left( n\right) $ is a stationary function. However, as was pointed out
above this function depends on time and, therefore, the period $T$ should
not be too big (it should be smaller than a characteristic time of variation
of histograms). Thus, to approach the instant picture one has to minimize
the possible value $T$ which reduces the number of points on the histogram
(usually in Ref. \cite{shn98} $N\sim 60-100$). The small value of $N$ causes
strong statistical fluctuations in the shape of histograms (typically $%
\Delta N^{2}\left( n\right) \sim 1/N\left( n\right) $) and, therefore, one
has to distinguish somehow the fine structure and fluctuations. In Ref. \cite
{shn98} this problem was solved as follows. Consider two signals $n\left(
t\right) $ and $m\left( t\right) $ from independent radioactive samples and
construct two series of histograms $P_{j}\left( n\right) $ and $W_{j}\left(
m\right) $. The fact that this signals are statistically independent means
that the standard analysis will show the absence of correlations between the
two signals. However, one can compare shapes of histograms $P_{j}\left(
n\right) $ and $W_{j^{\prime }}\left( m\right) $ (upon the transformation to
the normal variables $x=\left( n-<n>\right) /\sqrt{<n>}$) and build the
function $\nu \left( k\right) $ which is the number of coincidences as a
function of $k=j-j^{\prime }$. In the case of infinite series, if there are
only ordinary statistical fluctuations (caused by insufficient statistics),
one should find that $\nu $ does not depend on $k$ at all (finite series
with a length $L$ produce the dependence on $k$ as follows: $\nu _{0}\left(
k\right) =\nu _{0}L/\left( L-k\right) $) and indeed, computer simulations of
a random process give $\nu =\nu _{0}=const$ \cite{shn}. However,
radioactivity processes were shown to produce a function $\nu \left(
k\right) $ with a strong peak at $k=0$, with $\nu \left( 0\right) \gg \nu
_{0}$ (there also exists the so-called nearest zone $\nu \left( 1\right) \gg
\nu _{0}$) \cite{shn98}.

Unfortunately, this method speaks nothing about the fine structure itself
but shows the existence of a correlation between shapes of simultaneous
histograms. The most subtle moment here is the method used for comparison of
shapes of histograms and just this point causes main debates \cite{der2000}.
It is important, however, that the method was tested on real random signals
(computer simulations) which makes me think that the results are credible.

The correct way to solve all doubts should be the direct measurement of the
instant picture. This means that one should prepare a sufficiently big
number of identical samples, which will produce independent random
quantities $n_{i}$ ($i=1,...M$), and obtain $M$ simultaneous signals $%
n_{i}\left( t\right) $. Such a measurement will directly produce the
probability distribution as a function of time $P_{t}\left( n\right) =\frac{1%
}{M}\sum_{i}\delta \left( n-n_{i}\left( t\right) \right) $. On practice it
may be rather difficult to increase the number of simultaneous experiments
and one will probably have to combine both methods. Nevertheless, we stress,
that this is the only direct way to investigate the instant probability
density distribution. In fact, in the case of radioactivity it is possible
in principle to prepare a sufficiently big ($\sim 10^{3}$) number of equal
samples to take the instant picture.

Now consider the mechanism which can impose a fine structure on the Poisson
distribution in radioactivity processes. There exists a universal (though
rather trivial) mechanism which can be applied to physical processes of
diverse nature. It consist in the fact that our Universe (being a classical
system) can be in a quantum state which represents a mixture of two
different topologies.

Topology changes are expected to take place only at Planck scales and cause
an essential interest in quantum gravity \cite{wheeler,wormholes,tch}. In
particular, such processes are known to cause the loss of quantum coherence
and should be suppressed now, at least, there exist very strong experimental
restrictions, which come from oscillation experiments ($K\overline{K}$ and $%
\nu _{\mu }\leftrightarrow \nu _{\tau }$ oscillations), e.g., see, Ref. \cite
{osc} and references therein. This means that in the present Universe
topology changes should have a virtual character (the spacetime foam) and do
not affect (at least directly) all the standard physical processes. We may
expect, however, that in the early universe (during the quantum stage of the
evolution) such processes were real. After the quantum period, processes
with topology variations are suppressed and the topological structure of
space has to be preserved. Thus, we may hope that remnants of such processes
do survive till our days and may somehow display themselves (possible
effects of this sort were discussed in Ref. \cite{k99}). Moreover, due to
the expansion of the Universe a nontrivial topological structure should
display itself in the first place at macroscopic scales \cite{k99}. Thus, if
a portion of space was initially in a quantum state which mixes two
topologies, then now the topology of that region should not be fixed. This
hope is supported by the presumption that there is no a natural physical
process which is able to enforce the Universe to distinguish, on the
quasiclassical stage, one particular topology, if the initial quantum state
was a mixture of different topologies.

Let a portion of our space $V$ be in a quantum state which mixes two
topologies. The simplest state of this kind is described by the wave
function of the type 
\begin{equation}
\left| \Psi _{V}\right\rangle =a\left| V\right\rangle +b\left|
V^{+},V^{-}\right\rangle ,  \label{wf}
\end{equation}
where $\left| V\right\rangle $ represents the simple topology state and the
state $\left| V^{+},V^{-}\right\rangle $ describes two copies of the region $%
V$ ($V^{+}$ and $V^{-}$). In what follows, the origin of such a state is not
important. In the third quantization picture \cite{thq} this state can be
considered as if there is a small probability $\left| b\right| ^{2}$ that
the Universe has two copies. Schematically this can be illustrated by the
graph 
\begin{equation}
\frac{\;\;\;\;\;\;\;\;\;\;\;}{\;\;\;\;\;\;\;\;\;\;}\;{\bf =\;}\frac{\,}{\;\;}%
\frac{\;}{\;\;}\frac{\;}{\;\;}\frac{\;}{\;\;}\;{\bf +\;}_{\frac{\;}{\;\;}%
\frac{\;}{\;\;}\frac{\;}{\;\;}\frac{\;}{\;\;}\frac{\;}{\;\;}}^{\frac{\;}{\;\;%
}\frac{\;}{\;\;}\frac{\;}{\;\;}\frac{\;}{\;\;}\frac{\;}{\;\;}}{\bf \;.}
\label{gr1}
\end{equation}
From the other hand side, such a state appears if the region $V$ cuts a
portion of the space with a nontrivial topology, e.g., a gigantic wormhole,
which can be illustrated by the graph 
\begin{equation}
\frac{\;\;\;\;\;\;\;\;\;\;\;}{\;\;\;\;\;\;\;\;\;\;}\;{\bf =\;}\frac{\;}{\;\;}%
\frac{\;}{\;\;}\frac{\;}{\;\;}\frac{\;}{\;\;}{\bf \;+\;\frac{\;}{\;\;}\frac{%
\;}{\;\;}\langle } 
\begin{array}{c}
\frac{\;}{\;\;}\frac{\;}{\;\;}\frac{\;}{\;\;} \\ 
\frac{\;}{\;\;}\frac{\;}{\;\;}\frac{\;}{\;\;}
\end{array}
\rangle {\bf \frac{\;}{\;\;}\frac{\;}{\;\;}\;.}  \label{gr2}
\end{equation}
Moreover, in the case when the scale of the wormhole exceeds the horizon
size, both cases are indistinguishable for an observer who lives in the
middle of the wormhole. The normalization condition gives $\left| a\right|
^{2}+\left| b\right| ^{2}=1$ (we suppose the states $\left| V\right\rangle $
and $\left| V^{+},V^{-}\right\rangle $ to be normalized and orthogonal $%
\left\langle V|V^{+},V^{-}\right\rangle =0$).

The important question which arises here is how this state may be in
agreement with the classical nature of space and what an observer will see
living in the space of this sort. The first question has a simple answer. It
is the fact that properties of all three regions $V$, $V^{+}$, and $V^{-}$
should be very close to each other (matter distribution, etc.). Otherwise
there will appear a rather big minimal inevitable uncertainty in all
observable physical quantities. However small difference may (and probably
must) exist. To answer the second question we note that there exist at least
two approaches. First one comes from the third quantization scheme \cite{thq}
in which the state (\ref{wf}) means that the observer will be either in a
single Universe $V$, or in one of the two universes $V^{+}$ or $V^{-}$. It
is important, however, that both universes $V^{+}$ and $V^{-}$ are
indistinguishable, so the observer is not able to know exactly which one he
belongs to and, therefore, any observation will mix information about both
universes. We may say that the observer lives in both universes
simultaneously. However a portion of information about the state is lost and
this leads, in the general case, to the loss of quantum coherence widely
discussed in Refs. \cite{wormholes,tch} (see also discussions in Refs. \cite
{dec}). We note, however, that in this picture the loss of coherence is not
a dynamical process, for there is no a real change of the topology of space.
The dynamical change took place in the early Universe (but not now) and the
loss of the coherence happened then. Rather, in such a space pure states are
not reachable in principle and all physical objects are described by a
density matrix from the very beginning.

The second approach was proposed in Ref. \cite{k99}. In this approach the
state with two universes will effectively look as if all observables acquire
a double-valued character. In this case there also holds the identity
principle for both values of every observable and a part of information
about the quantum state also can be lost. This depends already on the nature
of measurements, e.g., if we measure the number of particles we will loose
the information about the double-valued nature of fields \cite{k99}.
However, in principle, there may exist a detector able to distinguish
double-valued and single valued quantities and, thus, to read all the
information about the quantum state.

For the sake of simplicity in what follows we discuss the third quantization
approach, though the same consideration (with minor modification) remains
valid in the second approach too.

Consider a quasiclassical system which is in the region $V$. Since the
structure of the space is not defined, the evolution of the system cannot be
described from the classical standpoint and, therefore, dynamics of the
system has to be described by a wave function. If the space had a simple
topology, the system would be described by a wave function of the form \cite
{Landau} 
\begin{equation}
\Psi _{1}\left( q\right) =A_{1}\left( q\right) e^{iS\left( q\right) },
\end{equation}
where $S\left( q\right) $ is the classical action, $q$ is a configuration
variable describing the system, and $A$ is a slow function of $q$ such that $%
P_{1}\left( q\right) =A_{1}^{2}(q)$ gives the probability distribution for
the observable $q.$ One usually supposes that this distribution is the
ordinary Gaussian distribution around an average value $q_{cl}=\left\langle
q\right\rangle $. The value $q_{cl}$ traces the classical trajectory which
can be found from the Hamilton-Jacobi equations 
\begin{equation}
\frac{\partial S}{\partial t}=-H\left( q,p\right) ,\;\;\;p=\frac{\partial S}{%
\partial q},  \label{eq}
\end{equation}
Here $S\left( q\right) $ is the action functional for the system and $%
H\left( p,q\right) $ is the appropriate Hamiltonian. In the case of the
state (\ref{wf}) the wave function acquires the form 
\begin{equation}
\Psi =a\Psi _{1}\left( q\right) +b\Psi _{2}\left( q^{+},q^{-}\right) 
\end{equation}
and contains one extra variable. Thus, to get the probability distribution
for the variable $q$ we have to integrate out the extra variable 
\begin{equation}
P\left( q\right) =\left| a\right| ^{2}P_{1}\left( q\right) +\left| b\right|
^{2}P_{2}\left( q\right) =\left| a\right| ^{2}\left| \Psi _{1}\left(
q\right) \right| ^{2}+\left| b\right| ^{2}\int \left| \Psi _{2}\left(
q,y\right) \right| ^{2}dy  \label{prb}
\end{equation}
and the quantum state of the system will be described by a density matrix.
In general this results in the loss of quantum coherence discussed in Refs. 
\cite{wormholes,osc}. It is important that the degree of the loss of
information depends essentially on the choice of the quantum state and it
can be shown that oscillation experiments \cite{osc} do not impose severe
restrictions on the choice of that state. It turns out that the state (\ref
{wf}) leads, in general, to spontaneous origin of an interference picture.

Indeed, the identity principle for the two regions $V^{+}$ and $V^{-}$
implies that the function $\Psi _{2}$ has the property 
\begin{equation}
\Psi _{2}\left( q^{+},q^{-}\right) =\pm \Psi _{2}\left( q^{-},q^{+}\right) ,
\end{equation}
where the sign $\pm $ depends on the choice of the statistics (Bose or Fermi
statistics). The quasiclassical nature of the system gives 
\begin{equation}
\Psi _{2}\left( q^{+},q^{-}\right) =\frac{1}{r}\left( \Psi _{+}\left(
q^{+}\right) \Psi _{-}\left( q^{-}\right) \pm \Psi _{+}\left( q^{-}\right)
\Psi _{-}\left( q^{+}\right) \right) ,  \label{2}
\end{equation}
where $r$ is the normalization constant and ($\alpha ,\beta =\pm $) 
\begin{equation}
\Psi _{\alpha }\left( q^{\beta }\right) =A_{\alpha }\left( q^{\beta }\right)
\exp \left( iS_{\alpha }\left( q^{\beta }\right) \right) .  \label{3}
\end{equation}
We suppose that these functions are normalized $\int dq\left| \Psi _{\alpha
}\right| ^{2}=1$. Then from (\ref{prb})-(\ref{3}) we get 
\begin{equation}
P_{2}\left( q\right) =\frac{1}{r^{2}}\left( P_{+}+P_{-}\pm 2f\sqrt{P_{+}P_{-}%
}\cos \left( \Delta S-\Delta \right) \right) ,
\end{equation}
where $P_{\pm }=A_{\pm }^{2}$, $\Delta S=S_{+}-S_{-}$, and constants $f$ and 
$\Delta $ are determined as follows 
\begin{equation}
fe^{i\Delta }=\int \Psi _{+}^{\ast }\left( q\right) \Psi _{-}\left( q\right)
dq.
\end{equation}
The normalization condition gives the relation $r^{2}=2\left( 1\pm
f^{2}\right) $. As was pointed out above, physical properties of the three
regions $V$, $V^{+}$, and $V^{-}$ should be very close to each other. This
means that all the functions $A\left( q\right) $, $A_{+}\left( q\right) $,
and $A_{-}\left( q\right) $ are very close and in the first approximation we
can put them to be equal $A\left( q\right) =$ $A_{\pm }\left( q\right) $ (it
is so if the difference of classical values $\delta q_{cl}^{\pm
}=q_{cl}-q_{cl}^{\pm }$\ is sufficiently small $\delta q_{cl}\ll
\left\langle \Delta q\right\rangle $). The same is also correct with respect
to the phases $S_{\pm }$ and we can use the decomposition 
\begin{equation}
\Delta S=\delta p_{cl}q+\theta \left( t\right) .
\end{equation}
Thus, we find (here $\varphi =\theta -\Delta $) 
\begin{equation}
P_{2}\left( q\right) \approx P_{1}\left( q\right) \frac{1}{1\pm f^{2}}\left(
1\pm f\cos \left( \delta p_{cl}q+\varphi \right) \right)
\end{equation}
and for the total distribution we get 
\begin{equation}
P\left( q\right) \approx P_{1}\left( q\right) \left( C_{1}+C_{2}\cos \left(
\delta p_{cl}\;q+\varphi \right) \right) .  \label{p}
\end{equation}
where 
\begin{equation}
C_{1}=\left| a\right| ^{2}+\left| b\right| ^{2}\frac{1}{1\pm f^{2}}%
,\;\;C_{2}=\pm \left| b\right| ^{2}\frac{f}{1\pm f^{2}}.
\end{equation}
It is important to note that in general the phase $\varphi $ and the value $%
q_{cl}\left( t\right) =\left\langle q\right\rangle $ depend on time $t$ and
so does the probability distribution $P\left( q\right) .$ In simplest cases
this dependence can be considered as a periodic function with some period $T$%
. Then, after averaging over the period $T$ we get already a stationary
distribution $\overline{P\left( x\right) }_{T}=P_{1}\left( x\right) $ (where 
$x=q-q_{cl}$) which reduces to the single Universe case.

We note that in (\ref{p}) the set of parameters $\delta p_{cl}$, $\varphi $,
and those which define $P_{1}\left( q\right) $ are specific parameters for
any given quasiclassical system. On the contrary, the parameter $\left|
b\right| ^{2}$ which defines the depth of modulation of the standard
distribution $P_{1}$ should represent a common for all physical systems
parameter which reflects the quantum structure of the region $V$.

Consider now a set of radioactive atoms. In the simplest case this set can
be well-modeled by a two-level system. Then it is easy to find that the wave
function takes the form 
\begin{equation}
\Psi \left( n\right) =\sqrt{\frac{N!}{n!\left( N-n\right) !}}\left( \sin
\alpha \right) ^{n}\left( \cos \alpha \right) ^{N-n}e^{inp-iEt},
\end{equation}
where $N$ is the total number of radioactive atoms, $n$ is the number of
fissions, $\sin ^{2}\alpha =\Gamma \Delta t=\frac{m\left( \Delta t\right) }{N%
}$ is the probability of a fission (in general $\Gamma $ depends on the
energy of the products of the fission), $m\left( t\right) =\left\langle
n\right\rangle $ is the mean number of fissions after the time $t$, $E$ is
the total energy of the system and $p$ is a phase parameter which is defined
by the tunneling process. In real experiments the time interval $\Delta t$
is such that the mean number of fissions is large enough, $\ m\gg 1$ and,
therefore, with very good approximation one gets ($\alpha =1,\pm $) 
\begin{equation}
A_{\alpha }\left( n\right) =\left( 2m_{\alpha }\pi \right) ^{-1/4}\exp
\left( -\frac{\left( n-m_{\alpha }\right) ^{2}}{4m_{\alpha }}\right)
,\;S_{\alpha }=p_{\alpha }n-E_{\alpha }t.  \label{pr}
\end{equation}
Thus, the existence of a small difference in the parameters $p_{\pm }$, $%
E_{\pm }$, which may exist in regions $V_{\pm }$, will impose a fine
structure (modulation) on the standard Poisson distribution which changes
with a characteristic period $T\sim \left( \delta E-\delta p\frac{d}{dt}%
m\right) ^{-1}$ (here $\delta E=E_{+}-E_{-}$ and we suppose that $m_{\alpha
}=m$).

The main aim of this letter was to present one of possibilities of the
origin of the fine structure in probability distributions, while the further
investigation of the problem discussed requires, in the first place, the
experimental confirmation of the effect by at least independent experimental
groups. We also note, that if such an effect really does exist, it may
become a principally new probe for the structure of our space.

\bigskip

I am grateful to S.E. Shnoll and D. Turaev for valuable discussions, and M.
Rainer for hospitality in Potsdam University where the main part of this
work was done. This research was supported by RFBR (Grant No. 98-02-16273)
and DFG (Grant No. 436 Rus 113/23618).


\end{document}